# MULTIPLE INTRABEAM SCATTERING IN X-Y COUPLED FOCUSING SYSTEMS

Valeri Lebedev[1] and Sergei Nagaitsev[2]

Fermilab, PO Box 500, Batavia, IL 60510-5011

*Abstract*

This paper describes an analytical method to calculate the emittance growth rates due to intra-beam scattering in a circular accelerator with arbitrary *x-y* coupling. The underlying theory is based on the Landau collision integral and the extended Mais-Ripken parametrization of a coupled betatron motion. The presented results are based on calculations of average emittance growth rates for an initially Gaussian distribution. They are applicable to both bunched and continuous beams.

## 1. Introduction

Intra-beam scattering (IBS) of charged particles in a beam results in the exchange of energy between various degrees of freedom, resulting in the increase of the average energy of particles in the beam frame and, generally, of the total beam emittance in the 6D phase space. The total Coulomb cross section of a two-particle scattering process in a vacuum diverges. However, in a plasma (or in a beam) it has a finite value due to field screening by other particles [1] or finite beam dimensions. Usually, two scattering regimes are considered: (1) single scattering, when a rare single collision produces a large change of particle momentum (the Touschek effect), and (2) multiple scattering, when multiple small angle frequent collisions cause diffusion. The former is usually responsible for the creation of distribution tails and the beam loss in electron machines, while the latter for changes in the distribution core.

IBS in accelerators is already a rather well-understood subject. The first decisive published work appears to be that of Piwinski [2], followed by Bjorken and Mtingwa [3]. These two earlier publications were both carried out from the first principles of two-body Coulomb collisions and largely ignored prior works on multiple scattering in a plasma [4, 5]. Ref. [6] utilized an approach based on the Landau kinetic equation [4], and gave the results, identical to [3]. In the present paper, following the same approach, we develop the IBS theory, which may be used in the case of arbitrary *x-y* coupling and is applicable to all circular accelerators. If required, the proposed method may be easily extended to a more general case of 3D coupling. Such an extension makes the formulas more complicated and is not presently needed for any existing storage rings because of their small synchrotron frequency values. Therefore, we limit our consideration to *x-y* coupling only. Like Ref. [6], the theoretical results include closed-form IBS rate expressions for beams with an arbitrary coupled betatron motion in the presence of both the vertical and horizontal dispersions. The results are presented in a matrix form and use symmetric elliptic integrals [7]. In this paper we are using a right-handed coordinate system and assume non-relativistic particle motion in the beam frame (BF). We also assume that the velocity spread in the BF is sufficiently large so that a usage of the plasma perturbation theory would be justified (see below). IBS growth rates with 3D coupling have been previously considered in Ref. [8], where the authors assume the knowledge of the so-called local momentum matrix, but do not give explicit expressions of how to obtain such a matrix. In this paper, we use the extended Mais-Ripken parametrization and derive the local momentum distribution explicitly, using the coupled-optics Twiss parameters.

---

[1] Also at JINR, Dubna, Moscow Reg., 141980, Russia

[2] Also at the University of Chicago, Chicago, IL 60637, USA

In the first part of this paper, we show how to derive growth rates for a single-component nonrelativistic plasma using the Landau collision integral. The novel result in this section is that we separate the diffusion and the friction terms and show them explicitly. In the second part of this paper, we apply the Landau collision integral approach to a relativistic particle beam in a laboratory frame and calculate the emittance growth rates.

## 2. Multiple Intrabeam Scattering in a single component plasma

An evolution of the particle velocity ($\mathbf{v}$) distribution function, $f \equiv f(\mathbf{v})$, due to multiple intrabeam scattering (IBS) in a uniform-density single component non-relativistic plasma, is described by the Landau collision integral [4],

$$\frac{\partial f}{\partial t} = -\frac{2\pi e^4 n L_c}{m^2}\frac{\partial}{\partial \mathrm{v}_i}\int\left( f\frac{\partial f'}{\partial \mathrm{v}'_j} - f'\frac{\partial f}{\partial \mathrm{v}_j}\right)\frac{u^2\delta_{ij} - u_i u_j}{u^3}d^3\mathrm{v}', \quad \mathbf{u} = \mathbf{v} - \mathbf{v}', \quad \int f d^3\mathrm{v} = 1, \tag{1}$$

where $n$ is the plasma density, $e$ and $m$ are the particle charge and mass, $L_c$ is the Coulomb logarithm, $f' \equiv f(\mathbf{v}')$, and $\delta_{ij}$ is the Kronecker delta. The indices $i, j$ = 1, 2, 3 denote $x$, $y$, $z$ axes, and the repeated indices are implicitly summed over. For clarity, we will re-write it in the following form:

$$\frac{\partial f}{\partial t}=\frac{\partial}{\partial \mathrm{v}_i}\left(-\frac{F_i}{m}f+\frac{D_{ij}}{m^2}\frac{\partial f}{\partial \mathrm{v}_j}\right) , \tag{2}$$

where the friction force and the diffusion are:

$$\begin{aligned}
F_i&=\frac{2\pi ne^4L_c}{m}\int\frac{\partial f'}{\partial \mathrm{v}'_j}\frac{u^2\delta_{ij}-u_iu_j}{u^3}d^3\mathrm{v}'=-\frac{4\pi ne^4L_c}{m}\int f'\frac{u_i}{u^3}d^3\mathrm{v}' ,\\
D_{ij}&=2\pi ne^4L_c\int f'\frac{u^2\delta_{ij}-u_iu_j}{u^3}d^3\mathrm{v}' ,
\end{aligned} \tag{3}$$

To obtain the second equality in the friction force of Eq. (3) we used integration by parts and accounted for

$$\frac{\partial}{\partial \mathrm{v}'_i}\left(\frac{u^2\delta_{ij}-u_iu_j}{u^3}\right)=2\frac{u_i}{u^3} . \tag{4}$$

To proceed further, we first consider how the average and rms velocities of a single particle are changing in time. To achieve this objective, we fix the diffusion and the friction force and set the initial distribution to a delta-function: $f=\delta\left(\mathrm{v}-\mathrm{v}_0\right)$. Then for the average particle velocity we obtain:

$$\begin{aligned}
\frac{d}{dt}\overline{\mathrm{v}}_i&=\int \mathrm{v}_i\frac{\partial f}{\partial t}\mathrm{d}^3\mathrm{v}=\int \mathrm{v}_i\frac{\partial}{\partial \mathrm{v}_l}\left(-\frac{F_l}{m}f+\frac{D_{kl}}{m^2}\frac{\partial f}{\partial \mathrm{v}_k}\right)\mathrm{d}^3\mathrm{v}=-\int\delta_{il}\left(-\frac{F_l}{m}f+\frac{D_{kl}}{m^2}\frac{\partial f}{\partial \mathrm{v}_k}\right)\mathrm{d}^3\mathrm{v}\\
&=\int\left(\frac{F_i}{m}f-\frac{D_{ki}}{m^2}\frac{\partial f}{\partial \mathrm{v}_i}\right)\mathrm{d}^3\mathrm{v}=\int\left(\frac{F_i}{m}+\frac{1}{m^2}\frac{\partial D_{ki}}{\partial \mathrm{v}_i}\right)f\,\mathrm{d}^3\mathrm{v}\xrightarrow{f=\delta(\mathbf{v}-\mathbf{v}_0)}\left.\left(\frac{F_i}{m}+\frac{1}{m^2}\frac{\partial D_{ki}}{\partial \mathrm{v}_i}\right)\right|_{\mathbf{v}=\mathbf{v}_0}
\end{aligned} \tag{5}$$

where we used Eq. (2), then performed integration by parts with $d\mathrm{v}_i/d\mathrm{v}_j=\delta_{ij}$. Similar for the rms velocity change we obtain:

$$\begin{aligned}
\frac{d}{dt}\overline{\delta \mathrm{v}_i\delta \mathrm{v}_j}&\equiv\frac{d}{dt}\left(\overline{(\mathrm{v}_i\text{-}\mathrm{v}_{0i})(\mathrm{v}_j\text{-}\mathrm{v}_{0j})}\right)=\int(\mathrm{v}_i\text{-}\mathrm{v}_{0i})(\mathrm{v}_j\text{-}\mathrm{v}_{0j})\frac{\partial}{\partial \mathrm{v}_l}\left(-\frac{F_l}{m}f+\frac{D_{kl}}{m^2}\frac{\partial f}{\partial \mathrm{v}_k}\right)\mathrm{d}^3\mathrm{v}\\
&=\int\left(\delta_{il}(\mathrm{v}_j\text{-}\mathrm{v}_{0j})+\delta_{jl}(\mathrm{v}_i\text{-}\mathrm{v}_{0i})\right)\left(\frac{F_l}{m}f-\frac{D_{kl}}{m^2}\frac{\partial f}{\partial \mathrm{v}_k}\right)\mathrm{d}^3\mathrm{v}=\int\left(\frac{F_i}{m}(\mathrm{v}_j\text{-}\mathrm{v}_{0j})+\frac{F_j}{m}(\mathrm{v}_i\text{-}\mathrm{v}_{0i})\right)f\,\mathrm{d}^3\mathrm{v}\\
&+\int\left(\left(\delta_{il}\delta_{jk}+\delta_{jl}\delta_{ik}\right)\frac{D_{kl}}{m^2}+\left(\delta_{il}(\mathrm{v}_j\text{-}\mathrm{v}_{0j})+\delta_{jl}(\mathrm{v}_i\text{-}\mathrm{v}_{0i})\right)\frac{1}{m^2}\frac{\partial D_{kl}}{\partial \mathrm{v}_k}\right)f\mathrm{d}^3\mathrm{v}\\
&\xrightarrow{f=\delta(\mathbf{v}-\mathbf{v}_0)}\int\frac{2D_{ij}}{m^2}f\mathrm{d}^3\mathrm{v}=\left.\frac{2D_{ij}}{m^2}\right|_{\mathbf{v}=\mathbf{v}_0} ,
\end{aligned} \tag{6}$$

where we took advantage of $D_{ij}$ being a symmetric tensor. As one can see from Eq. (5), the dependence of diffusion on the velocity results in an addition to the cooling force so that the effective cooling force in a single component plasma is:

$$\tilde{F}_i = F_i + \frac{1}{m}\frac{\partial D_{ki}}{\partial \mathrm{v}_i} \quad . \tag{7}$$

Note that the second term is formally present in the cooling force of electron cooling [9], but its value is suppressed by the ratio of proton to electron masses and therefore is generally ignored.

Now, we derive the contribution of the "actual diffusion" described by Eq. (6) to the particle diffusion for the case of Gaussian distribution with three different temperatures:

$$f = \frac{1}{(2\pi)^{3/2}\sigma_{\mathrm{v}x}\sigma_{\mathrm{v}y}\sigma_{\mathrm{v}z}}\exp\left(-\frac{1}{2}\left(\frac{\mathrm{v}_x^2}{\sigma_{\mathrm{v}x}^2}+\frac{\mathrm{v}_y^2}{\sigma_{\mathrm{v}y}^2}+\frac{\mathrm{v}_z^2}{\sigma_{\mathrm{v}z}^2}\right)\right) . \tag{8}$$

Here $\sigma_{\mathrm{v}x}$, $\sigma_{\mathrm{v}y}$, and $\sigma_{\mathrm{v}z}$, are the rms velocities for the corresponding degrees of freedom and we choose the coordinate frame along the main axes of the velocity ellipsoid. Such a choice of the coordinate frame diagonalizes the diffusion tensor, consequently, we need to compute the diagonal terms only.

Considering that for the diffusion contribution $\overline{\mathrm{v}_i \delta \mathrm{v}_j} = 0$ we obtain that $d\left(\overline{\mathrm{v}_i^{\,2}}\right)/dt\Big|_D =$ $d\left(\overline{\delta \mathrm{v}_i^{\,2}}\right)/dt\Big|_D$. Then, using Eq. (6) we obtain the time derivative of the rms velocity change in a plasma due to diffusion, averaged over all particles. For the $x$-coordinate we have:

$$\begin{aligned}\frac{\partial}{\partial t}\overline{\mathrm{v}_x^{\,2}}\Big|_D &= \frac{2}{m^2}\int D_{11} f d^3\mathrm{v} = \frac{4\pi n e^4 L_c}{m^2}\int f f' \frac{u_y^{\,2}+u_z^{\,2}}{u^3} d^3\mathrm{v}' d^3\mathrm{v} \\ &= \frac{4\pi n e^4 L_c}{m^2 (2\pi)^3 \sigma_{\mathrm{v}x}^{\,2}\sigma_{\mathrm{v}y}^{\,2}\sigma_{\mathrm{v}z}^{\,2}}\int \mathrm{e}^{-\frac{\mathrm{v}_x^{\,2}}{2\sigma_{\mathrm{v}x}^{\,2}}-\frac{\mathrm{v}_y^{\,2}}{2\sigma_{\mathrm{v}y}^{\,2}}-\frac{\mathrm{v}_z^{\,2}}{2\sigma_{\mathrm{v}z}^{\,2}}}\,\mathrm{e}^{-\frac{\mathrm{v}_x'^{\,2}}{2\sigma_{\mathrm{v}x}^{\,2}}-\frac{\mathrm{v}_y'^{\,2}}{2\sigma_{\mathrm{v}y}^{\,2}}-\frac{\mathrm{v}_z'^{\,2}}{2\sigma_{\mathrm{v}z}^{\,2}}}\frac{u_y^{\,2}+u_z^{\,2}}{u^3} d^3\mathrm{v}' d^3\mathrm{v} .\end{aligned} \tag{9}$$

Transition to the new variables $\mathbf{u} = \mathbf{v} - \mathbf{v}'$ and $\mathbf{w} = \mathbf{v} + \mathbf{v}'$ yields:

$$\begin{aligned}\frac{\partial}{\partial t}\overline{\mathrm{v}_x^{\,2}}\Big|_D &= \frac{4\pi n e^4 L_c}{8m^2 (2\pi)^3 \sigma_{\mathrm{v}x}^{\,2}\sigma_{\mathrm{v}y}^{\,2}\sigma_{\mathrm{v}z}^{\,2}}\int \mathrm{e}^{-\frac{u_x^{\,2}+w_x^{\,2}}{4\sigma_{\mathrm{v}x}^{\,2}}-\frac{u_y^{\,2}+w_y^{\,2}}{4\sigma_{\mathrm{v}y}^{\,2}}-\frac{u_z^{\,2}+w_z^{\,2}}{4\sigma_{\mathrm{v}z}^{\,2}}}\frac{u_y^{\,2}+u_z^{\,2}}{u^3} d^3u d^3w \\ &= \frac{n e^4 L_c}{2\sqrt{\pi} m^2 \sigma_{\mathrm{v}x}\sigma_{\mathrm{v}y}\sigma_{\mathrm{v}z}}\int \mathrm{e}^{-\frac{u_x^{\,2}}{4\sigma_{\mathrm{v}x}^{\,2}}-\frac{u_y^{\,2}}{4\sigma_{\mathrm{v}y}^{\,2}}-\frac{u_z^{\,2}}{4\sigma_{\mathrm{v}z}^{\,2}}}\frac{u_y^{\,2}+u_z^{\,2}}{u^3} d^3u .\end{aligned} \tag{10}$$

For further integration we use the following identity:

$$\frac{1}{\theta^3} = \frac{1}{4\sqrt{\pi}}\int_0^\infty \sqrt{\lambda}\, e^{-\lambda\theta^2/4} d\lambda \ , \tag{11}$$

which yields

$$\left.\frac{\partial}{\partial t}\overline{\mathrm{v}_x^{\ 2}}\right|_D = \frac{e^4 n L_c}{2\sqrt{\pi} m^2 \sigma_{\mathrm{v}x}\sigma_{\mathrm{v}y}\sigma_{\mathrm{v}z}} \int_0^\infty \frac{d\lambda}{4\sqrt{\pi}}\sqrt{\lambda}\int e^{-\left(u_x^{\ 2}+u_x^{\ 2}+u_x^{\ 2}\right)\lambda/4}\left(u_y^{\ 2}+u_z^{\ 2}\right)\mathrm{e}^{-\frac{u_x^{\ 2}}{4\sigma_{\mathrm{v}x}^{\ 2}}-\frac{u_y^{\ 2}}{4\sigma_{\mathrm{v}y}^{\ 2}}-\frac{u_z^{\ 2}}{4\sigma_{\mathrm{v}z}^{\ 2}}}\, d^3u$$

$$= \frac{2\sqrt{\pi} e^4 n L_c}{m^2 \sigma_{\mathrm{v}x}\sigma_{\mathrm{v}y}\sigma_{\mathrm{v}z}} \int_0^\infty \frac{\sqrt{\lambda}\, d\lambda}{\left(\frac{1}{\sigma_{\mathrm{v}x}^{\ 2}}+\lambda\right)^{1/2}\left(\frac{1}{\sigma_{\mathrm{v}y}^{\ 2}}+\lambda\right)^{1/2}\left(\frac{1}{\sigma_{\mathrm{v}z}^{\ 2}}+\lambda\right)^{1/2}} \left(\frac{1}{\frac{1}{\sigma_{\mathrm{v}y}^{\ 2}}+\lambda}+\frac{1}{\frac{1}{\sigma_{\mathrm{v}z}^{\ 2}}+\lambda}\right). \tag{12}$$

This integral can be expressed through the Carlson symmetric elliptical integral [7]:

$$\mathrm{R_D}\left(x,y,z\right) = \frac{3}{2}\int_0^\infty \frac{dt}{\sqrt{\left(t+x\right)\left(t+y\right)\left(t+z\right)^3}} \ . \tag{13}$$

Performing the transformations, one obtains for the diffusion contribution:

$$\frac{d}{dt}\begin{pmatrix}\sigma_{\mathrm{v}x}^{\ 2}\\ \sigma_{\mathrm{v}y}^{\ 2}\\ \sigma_{\mathrm{v}z}^{\ 2}\end{pmatrix}_D = \frac{\left(2\pi\right)^{3/2} e^4 n L_c}{m^2\sqrt{\sigma_{\mathrm{v}x}^2+\sigma_{\mathrm{v}y}^2+\sigma_{\mathrm{v}z}^2}}\begin{pmatrix}\Psi_D\left(\sigma_{\mathrm{v}x},\sigma_{\mathrm{v}y},\sigma_{\mathrm{v}z}\right)\\ \Psi_D\left(\sigma_{\mathrm{v}y},\sigma_{\mathrm{v}z},\sigma_{\mathrm{v}x}\right)\\ \Psi_D\left(\sigma_{\mathrm{v}z},\sigma_{\mathrm{v}x},\sigma_{\mathrm{v}y}\right)\end{pmatrix}, \tag{14}$$

where

$$\Psi_D\left(x,y,z\right) = \frac{\sqrt{2}r}{3\pi}\left(y^2\,\mathrm{R_D}\left(z^2,x^2,y^2\right)+z^2\,\mathrm{R_D}\left(x^2,y^2,z^2\right)\right), \quad r=\sqrt{x^2+y^2+z^2}. \tag{15}$$

The evaluation of time derivatives for the rms velocities, which accounts for both the diffusion and the friction terms, was carried out in Ref. [7]. The result is:

$$\frac{d}{dt}\begin{pmatrix}\sigma_{\mathrm{v}x}^{\ 2}\\ \sigma_{\mathrm{v}y}^{\ 2}\\ \sigma_{\mathrm{v}z}^{\ 2}\end{pmatrix} = \frac{\left(2\pi\right)^{3/2} e^4 n L_c}{m^2\sqrt{\sigma_{\mathrm{v}x}^2+\sigma_{\mathrm{v}y}^2+\sigma_{\mathrm{v}z}^2}}\begin{pmatrix}\Psi\left(\sigma_{\mathrm{v}x},\sigma_{\mathrm{v}y},\sigma_{\mathrm{v}z}\right)\\ \Psi\left(\sigma_{\mathrm{v}y},\sigma_{\mathrm{v}z},\sigma_{\mathrm{v}x}\right)\\ \Psi\left(\sigma_{\mathrm{v}z},\sigma_{\mathrm{v}x},\sigma_{\mathrm{v}y}\right)\end{pmatrix} \tag{16}$$

where

$$\Psi\left(x,y,z\right) = \frac{\sqrt{2}r}{3\pi}\left(y^2\,\mathrm{R_D}\left(z^2,x^2,y^2\right)+z^2\,\mathrm{R_D}\left(x^2,y^2,z^2\right)-2x^2\,\mathrm{R_D}\left(y^2,z^2,x^2\right)\right) \ . \tag{17}$$

To obtain the contribution of the friction force we subtract Eq. (14) from Eq. (16). It yields:

$$\frac{d}{dt}\begin{pmatrix}\sigma_{\mathrm{v}x}^{\ 2}\\ \sigma_{\mathrm{v}y}^{\ 2}\\ \sigma_{\mathrm{v}z}^{\ 2}\end{pmatrix}_F = -\frac{\left(2\pi\right)^{3/2} e^4 n L_c}{m^2\sqrt{\sigma_{\mathrm{v}x}^2+\sigma_{\mathrm{v}y}^2+\sigma_{\mathrm{v}z}^2}}\begin{pmatrix}\Psi_F\left(\sigma_{\mathrm{v}x},\sigma_{\mathrm{v}y},\sigma_{\mathrm{v}z}\right)\\ \Psi_F\left(\sigma_{\mathrm{v}y},\sigma_{\mathrm{v}z},\sigma_{\mathrm{v}x}\right)\\ \Psi_F\left(\sigma_{\mathrm{v}z},\sigma_{\mathrm{v}x},\sigma_{\mathrm{v}y}\right)\end{pmatrix}, \tag{18}$$

where

$$\Psi_F\left(x,y,z\right)=\frac{2\sqrt{2}r}{3\pi}x^2\,\mathrm{R_D}\left(y^2,z^2,x^2\right)\ . \tag{19}$$

Finally, we write Eqs. (14) and (16) in a vector form:

$$\left.\frac{d\mathbf{\Sigma}_\mathrm{v}}{dt}\right|_F=-\frac{\left(2\pi\right)^{3/2}nr_0{}^2c^4L_c}{\sqrt{\sigma_{\mathrm{v}x}^2+\sigma_\mathrm{vy}^2+\sigma_\mathrm{vz}^2}}\mathbf{\Psi}_F\left(\mathbf{\Sigma}_\mathrm{v}\right),$$
$$\mathbf{\Psi}_F\left(\mathbf{\Sigma}_\mathrm{v}\right)=\begin{pmatrix}\Psi_F\left(\sigma_\mathrm{vx},\sigma_\mathrm{vy},\sigma_\mathrm{vz}\right) & 0 & 0\\ 0 & \Psi_F\left(\sigma_\mathrm{vy},\sigma_\mathrm{vz},\sigma_\mathrm{vx}\right) & 0\\ 0 & 0 & \Psi_F\left(\sigma_\mathrm{vz},\sigma_\mathrm{vx},\sigma_\mathrm{vy}\right)\end{pmatrix}. \tag{20}$$

$$\left.\frac{d\mathbf{\Sigma}_\mathrm{v}}{dt}\right|_D=\frac{\left(2\pi\right)^{3/2}nr_0{}^2c^4L_c}{\sqrt{\sigma_{\mathrm{v}x}^2+\sigma_{\mathrm{v}y}^2+\sigma_{\mathrm{v}z}^2}}\mathbf{\Psi}_D\left(\mathbf{\Sigma}_\mathrm{v}\right),$$
$$\mathbf{\Psi}_D\left(\mathbf{\Sigma}_\mathrm{v}\right)=\begin{pmatrix}\Psi_D\left(\sigma_{\mathrm{v}x},\sigma_{\mathrm{v}y},\sigma_{\mathrm{v}z}\right) & 0 & 0\\ 0 & \Psi_D\left(\sigma_{\mathrm{v}y},\sigma_{\mathrm{v}z},\sigma_{\mathrm{v}x}\right) & 0\\ 0 & 0 & \Psi_D\left(\sigma_{\mathrm{v}z},\sigma_{\mathrm{v}x},\sigma_{\mathrm{v}y}\right)\end{pmatrix} \tag{21}$$

Here $\mathbf{\Sigma}_\mathrm{v}=\mathrm{diag}\left(\sigma_{vx}^2,\sigma_{vy}^2,\sigma_{vz}^2\right)$ is a diagonal 3x3 matrix, $r_0=e^2/mc^2$ is the particle classical radius, and $c$ is the speed of light. For further consideration we need an explicit expression for the Coulomb logarithm: $L_c=\ln(\rho_\mathrm{max}/\rho_\mathrm{min})$, where

$$\begin{aligned}\rho_\mathrm{min}&=r_0c^2/\overline{\mathrm{v}^2}\ ,\\ \rho_\mathrm{max}&=\sqrt{\overline{\mathrm{v}^2}/4\pi nr_0c^2}\ ,\end{aligned}\qquad \overline{\mathrm{v}^2}=\sigma_{\mathrm{v}x}^2+\sigma_\mathrm{vy}^2+\sigma_\mathrm{vz}^2\quad . \tag{22}$$

Note that the considered inhere theory is applicable only if $L_c \gg 1$. This condition is not satisfied for deeply cooled beams, where potential (correlational) energy of particle interactions becomes close or larger than the average kinetic energy of particles.

An algorithm for fast numerical evaluations of $\mathrm{R_D}(u,v,w)$ is discussed in Ref. [7]. The functions $\Psi\left(x,y,z\right)=\Psi_D\left(x,y,z\right)-\Psi_F\left(x,y,z\right)$, $\Psi_F\left(x,y,z\right)$ and $\Psi_D\left(x,y,z\right)$ are chosen such that they depend on the ratios of their variables but not on the value of $r=\sqrt{x^2+y^2+z^2}$. The functions are symmetric with respect to the variables $y$ and $z$, and the function $\Psi\left(x,y,z\right)$ is normalized such that $\Psi\left(0,1,1\right)=1$. Conservation of energy yields: $\Psi(x,y,z)+\Psi(y,z,x)+\Psi(z,x,y)=0$, $\Psi(1,0,1)=\Psi(1,1,0)=-1/2$, and that $\Psi(1,1,1)=0$ as expected in a thermal equilibrium.

## 3. Landau Collision Integral in the Laboratory Frame

Consider a coasting beam of particles in the laboratory frame of reference with an average energy $E$ and an average momentum $p$, circulating in a storage ring with a circumference $C$. Let us introduce $\beta = (pc)/E$ and $\gamma$, the usual Lorentz factors. We will proceed as follows: (1) we will introduce the lab frame distribution function and the corresponding mode emittances, then (2) we will re-write the Landau Collision integral in the lab frame, finally applying it to the emittance growth rates.

In our calculations we will be using the extended Mais-Ripken parameterization of the Twiss parameters [10], which parametrizes the eigen-vectors $\mathbf{v}_i$ $(i = 1,\ 2)$ of an $x$-$y$ coupled motion through the Twiss parameters as following:

$$
\begin{aligned}
\mathbf{v}_1 &= \left( \sqrt{\beta_{1x}}, -\frac{i(1-u)+\alpha_{1x}}{\sqrt{\beta_{1x}}}, \sqrt{\beta_{1y}} e^{i\nu_1}, -\frac{iu+\alpha_{1y}}{\sqrt{\beta_{1y}}} e^{i\nu_1} \right)^T, \\
\mathbf{v}_2 &= \left( \sqrt{\beta_{2x}} e^{i\nu_2}, -\frac{iu+\alpha_{2x}}{\sqrt{\beta_{2x}}} e^{i\nu_2}, \sqrt{\beta_{2y}}, -\frac{i(1-u)+\alpha_{2y}}{\sqrt{\beta_{2y}}} \right)^T,
\end{aligned}
\tag{23}
$$

where $\alpha_{ix}, \alpha_{iy}, \beta_{ix}, \beta_{iy}$ ($i$ = 1,2) are the generalized Twiss functions. Three other real-valued functions, $u$ and $\nu_{1,\,2}$, can be expressed in terms of the Twiss functions. To remind, for the uncoupled motion, one would set

$$
\begin{aligned}
\mathbf{v}_1 &= \left( \sqrt{\beta_x}, -\frac{i+\alpha_x}{\sqrt{\beta_x}}, 0, 0 \right)^T, \\
\mathbf{v}_2 &= \left( 0, 0, \sqrt{\beta_y}, -\frac{i+\alpha_y}{\sqrt{\beta_y}} \right)^T.
\end{aligned}
\tag{24}
$$

The eigen-vectors are normalized by the condition of symplectic orthogonality [10]:

$$
\begin{aligned}
&\mathbf{v}_k^+ \mathbf{U} \mathbf{v}_k = -2i, \\
&\mathbf{v}_k^+ \mathbf{U} \mathbf{v}_m = 0, \qquad \text{for } k \neq m, \\
&\mathbf{v}_k^T \mathbf{U} \mathbf{v}_m = 0,
\end{aligned}
\tag{25}
$$

where the symbol $^+$ denotes a transposed and a complex conjugated vector ($\mathbf{v}_k^+ = \mathbf{v}_k^{*T}$), $k, m = 1, 2$, and the unit 4x4 symplectic matrix is:

$$
\mathbf{U} = \begin{bmatrix} \mathbf{J} & \mathbf{0} \\ \mathbf{0} & \mathbf{J} \end{bmatrix}, \quad \mathbf{J} = \begin{pmatrix} 0 & 1 \\ -1 & 0 \end{pmatrix}.
\tag{26}
$$

We will also use a symplectic 4x4 matrix, built from the eigen vectors:

$$\mathbf{V} = \left[ \operatorname{Re}\mathbf{v}_1, -\operatorname{Im}\mathbf{v}_1, \operatorname{Re}\mathbf{v}_2, -\operatorname{Im}\mathbf{v}_2 \right] . \qquad (27)$$

The symplecticity condition determines that:

$$\mathbf{V}^T\mathbf{U}\mathbf{V} = \mathbf{U} \quad \Leftrightarrow \quad \mathbf{V}\mathbf{U}\mathbf{V}^T = \mathbf{U} . \qquad (28)$$

An equation, expressing the distribution function of the beam in the lab frame in the 4D transverse phase space, was derived in Ref. [10]: $f(\mathbf{x}) = \left(4\pi^2\varepsilon_1\varepsilon_2\right)^{-1}\exp\left(-\mathbf{x}^T\mathbf{\Xi}\mathbf{x}/2\right)$. Here, we additionally account for the contributions of the relative momentum deviation, $\theta_s \equiv \Delta p / p$, which yields:

$$f_0\left(\mathbf{x},\theta_s\right) = \frac{1}{4\pi^2\sqrt{2\pi}\varepsilon_1\varepsilon_2\sigma_p}\exp\left(-\frac{1}{2}\left(\left(\mathbf{x}-\mathbf{D}\theta_s\right)^T\mathbf{\Xi}\left(\mathbf{x}-\mathbf{D}\theta_s\right)+\frac{\theta_s^{\,2}}{2\sigma_p^{\,2}}\right)\right) . \qquad (29)$$

where $\mathbf{x} = \left[x, \quad \theta_x, \quad y, \quad \theta_y\right]^T$ is the vector of particle coordinates in the 4D phase space, $\varepsilon_1$ and $\varepsilon_2$ are the mode emittances [10] defined below, $\sigma_p$ is the rms momentum spread, and $\mathbf{D} = \left[D_x, \quad D'_x, \quad D_y, \quad D'_y\right]^T$ is the vector of dispersions and their derivatives. The matrix $\mathbf{\Xi}$ can be expressed through the matrix $\mathbf{V}$ and a diagonal matrix, built from the mode emittances [10], $\mathbf{\Xi}' = \operatorname{diag}\left(1/\varepsilon_1, 1/\varepsilon_1, 1/\varepsilon_2, 1/\varepsilon_2\right)$, such that:

$$\mathbf{\Xi} = \mathbf{U}\mathbf{V}\mathbf{\Xi}'\mathbf{V}^T\mathbf{U}^T . \qquad (30)$$

A direct representation of the matrix $\mathbf{\Xi}$ through emittances and generalized Twiss parameters can be found in the appendix of Ref [10]. The choice of the vector $\mathbf{x}$ above implies zero longitudinal magnetic field, which is common in most accelerator optics codes, presenting the Twiss parameters after an exit from a solenoid. If the longitudinal magnetic field is present, the vector $\mathbf{x}$ has to be constructed from the canonical momenta and, consequently, it has the following form:

$$\mathbf{x} = \left[x, \quad \theta_x - Ry/2, \quad y, \quad \theta_y + Rx/2\right]^T ,$$

where $R = eB_s / pc$, $B_s$ is the longitudinal magnetic field, and $p$ is the particle momentum. However, the beam transverse sizes and local velocity spreads do not change with the transition from a non-zero to a zero longitudinal magnetic field and vice versa. Therefore, in further calculations we imply zero longitudinal magnetic field, which does not affect the generality of the obtained results.

To account for the longitudinal momentum spread in further derivations we introduce

$$\mathbf{\Xi}_{tot}=\begin{bmatrix}\Xi_{11} & \Xi_{12} & \Xi_{13} & \Xi_{14} & \Xi_{15}\\ \Xi_{12} & \Xi_{22} & \Xi_{23} & \Xi_{24} & \Xi_{25}\\ \Xi_{13} & \Xi_{23} & \Xi_{33} & \Xi_{34} & \Xi_{35}\\ \Xi_{14} & \Xi_{24} & \Xi_{34} & \Xi_{44} & \Xi_{45}\\ \Xi_{15} & \Xi_{25} & \Xi_{35} & \Xi_{45} & \Xi_{55}\end{bmatrix},\quad \Xi_{55}=\frac{1}{\sigma_p{}^2}+\mathbf{D}^T\mathbf{\Xi}\mathbf{D}\,,\quad \begin{bmatrix}\Xi_{15}\\ \Xi_{25}\\ \Xi_{35}\\ \Xi_{45}\end{bmatrix}=\mathbf{\Xi}\mathbf{D}\ , \tag{31}$$

where the upper left 4 x 4 corner is given by the matrix $\mathbf{\Xi}$, Eq. (30), and the distribution function is $f_0 \propto \exp\left(-\mathbf{z}^T\mathbf{\Xi}_{tot}\mathbf{z}/2\right)$ with $\mathbf{z}=\left[x,\ \ \theta_x,\ \ y,\ \ \theta_y,\ \ \theta_s\right]^T$. Particle velocities in the BF can be expressed through their angles in the lab frame: $\mathrm{v}_\perp=\gamma\beta c\theta_\perp, \mathrm{v}_\parallel=\beta c\theta_\parallel$. To account for this, we introduce the following vectors $\hat{\mathbf{x}}=\left[x\ \ \theta_x\ \ y\ \ \theta_y\ \ \theta_s/\gamma\right]^T$, $\mathbf{X}=\left[x\ \ \ y\right]^T$ and $\hat{\boldsymbol{\theta}}=\left[\theta_x\ \ \ \theta_y\ \ \ \theta_s/\gamma\right]^T$. The latter one allows one to write particle velocities as $\beta c\gamma\hat{\boldsymbol{\theta}}$. Consequently, we rewrite matrix $\mathbf{\Xi}_{tot}$ as $\hat{\mathbf{\Xi}}$:

$$\hat{\mathbf{\Xi}}=\begin{bmatrix}\Xi_{11} & \Xi_{12} & \Xi_{13} & \Xi_{14} & \gamma\Xi_{15}\\ \Xi_{12} & \Xi_{22} & \Xi_{23} & \Xi_{24} & \gamma\Xi_{25}\\ \Xi_{13} & \Xi_{23} & \Xi_{33} & \Xi_{34} & \gamma\Xi_{35}\\ \Xi_{14} & \Xi_{24} & \Xi_{34} & \Xi_{44} & \gamma\Xi_{45}\\ \gamma\Xi_{15} & \gamma\Xi_{25} & \gamma\Xi_{35} & \gamma\Xi_{45} & \gamma^2\Xi_{55}\end{bmatrix}. \tag{32}$$

Then, the distribution function in the lab frame can be finally rewritten in the following form

$$f=\frac{\gamma}{4\pi^2\sqrt{2\pi}\varepsilon_1\varepsilon_2\sigma_p}\exp\left(-\frac{1}{2}\hat{\mathbf{x}}^T\hat{\mathbf{\Xi}}\hat{\mathbf{x}}\right). \tag{33}$$

Consequently, the distribution is normalized as $\int f\,d^2Xd^3\hat{\theta}=1$. Notice that this is a coasting-beam distribution function. The transition to the bunched beam is straightforward and will be considered later.

We consider now how to calculate the rms beam emittances $\varepsilon_1$, $\varepsilon_2$, and the momentum spread $\sigma_p$ for an arbitrary beam distribution. Let the vector of particle positions be expressed through single particle actions, $J_1$, $J_2$ and phases, $\psi_1$, $\psi_2$:

$$\mathbf{x}=\frac{1}{2}\left(\sqrt{2J_1}e^{i\psi_1}\mathbf{v}_1+\sqrt{2J_2}e^{i\psi_2}\mathbf{v}_2+CC\right)+\mathbf{D}\theta_s\ . \tag{34}$$

Note that for the uncoupled motion, Eqs. (24) would yield the well-known expressions:

$$x=\sqrt{2\beta_x J_x}\cos\mu_x + D_x\theta_s ,$$
$$\theta_x = -\sqrt{\frac{2J_x}{\beta_x}}\left(\sin\mu_x+\alpha_x\cos\mu_x\right)+D'_x\theta_s , \tag{35}$$

representing the uncoupled analog of Eq. (34) and similarly for $y$ and $\theta_y$.

To find single-particle transverse actions, $J_1$, $J_2$, we use the conditions of symplectic orthogonality, Eq. (25),

$$\begin{aligned}\mathbf{v}_1{}^+\mathbf{U}\mathbf{x} &= \frac{1}{2}\left(\sqrt{2J_1}e^{i\psi_1}\mathbf{v}_1{}^+\mathbf{U}\mathbf{v}_1+\sqrt{2J_2}e^{i\psi_2}\mathbf{v}_1{}^+\mathbf{U}\mathbf{v}_2+\mathbf{v}_1{}^+\mathbf{U}(CC)\right)+\mathbf{v}_1{}^+\mathbf{U}\mathbf{D}\theta_s \\ &= \frac{1}{2}\sqrt{2J_1}e^{i\psi_1}(-2i)+\mathbf{v}_1{}^+\mathbf{U}\mathbf{D}\theta_s = -i\sqrt{2J_1}e^{i\psi_1}+\mathbf{v}_1{}^+\mathbf{U}\mathbf{D}\theta_s\end{aligned} \quad . \tag{36}$$

Regrouping, we obtain $\mathbf{v}_1{}^+\mathbf{U}\left(\mathbf{x}-\mathbf{D}\theta_s\right)=-i\sqrt{2J_1}e^{i\psi_1}$. Then, multiplying by a complex conjugate we obtain the first single particle action:

$$J_1=\frac{1}{2}\left(\mathbf{x}^+-\mathbf{D}^+\theta_s\right)\mathbf{U}^T\mathbf{v}_1\mathbf{v}_1{}^+\mathbf{U}\left(\mathbf{x}-\mathbf{D}\theta_s\right) . \tag{37}$$

Considering that $\mathbf{U}^+\mathbf{v}_1\mathbf{v}_1{}^+\mathbf{U}$ is a Hermitian matrix and $\left(\mathbf{x}-\mathbf{D}\theta_s\right)$ is a real vector, we introduce the symmetric matrices,

$$\tilde{\mathbf{V}}_1=\mathrm{Re}\left(\mathbf{U}^T\mathbf{v}_1\mathbf{v}_1{}^+\mathbf{U}\right),\quad \tilde{\mathbf{V}}_2=\mathrm{Re}\left(\mathbf{U}^T\mathbf{v}_2\mathbf{v}_2{}^+\mathbf{U}\right) . \tag{38}$$

Particle actions can then be written as

$$J_1=\frac{1}{2}\left(\mathbf{x}^T-\mathbf{D}^T\theta_s\right)\tilde{\mathbf{V}}_1\left(\mathbf{x}-\mathbf{D}\theta_s\right),$$
$$J_2=\frac{1}{2}\left(\mathbf{x}^T-\mathbf{D}^T\theta_s\right)\tilde{\mathbf{V}}_2\left(\mathbf{x}-\mathbf{D}\theta_s\right). \tag{39}$$

Finally, we can write the rms eigen emittances as:

$$\varepsilon_1=\int J_1 f\, d^2Xd^3\hat{\theta}=\frac{1}{2}\int\left(\mathbf{x}^T-\gamma\mathbf{D}^T\hat{\theta}_s\right)\tilde{\mathbf{V}}_1\left(\mathbf{x}-\gamma\mathbf{D}\hat{\theta}_s\right)f\, d^2Xd^3\hat{\theta},$$
$$\varepsilon_2=\int J_2 f\, d^2Xd^3\hat{\theta}=\frac{1}{2}\int\left(\mathbf{x}^T-\gamma\mathbf{D}^T\hat{\theta}_s\right)\tilde{\mathbf{V}}_2\left(\mathbf{x}-\gamma\mathbf{D}\hat{\theta}_s\right)f\, d^2Xd^3\hat{\theta}, \tag{40}$$
$$\sigma_p{}^2=\gamma^2\int\hat{\theta}_s{}^2 f\, d^2Xd^3\hat{\theta} ,$$

where $\mathbf{X}=\left[x \quad y\right]^T$, $\hat{\boldsymbol{\theta}}=\left[\theta_x \quad \theta_y \quad \hat{\theta}_s\right]^T$, and $\hat{\theta}_s=\theta_s/\gamma$. Note that Eqs. (40) are valid for an arbitrary distribution function *f*. For a Gaussian distribution function (33), Eq. (40) gives three identity equations: $\varepsilon_1=\varepsilon_1$, $\varepsilon_2=\varepsilon_2$, and $\sigma_p=\sigma_p$.

One can notice that the particle actions (39) can be expressed as quadratic forms,

$$J_q = a_{ij}^{(q)}\hat{\theta}_i\hat{\theta}_j + b_{i\alpha}^{(q)}\hat{\theta}_i X_\alpha + c_{\alpha\beta}^{(q)} X_\alpha X_\beta\ ; \quad i,j=1,2,3; \quad \alpha,\beta,q=1,2\ . \tag{41}$$

Similarly, the distribution function, Eq. (33), is:

$$f = \frac{\gamma}{4\pi^2\sqrt{2\pi}\varepsilon_1\varepsilon_2\sigma_p}\exp\left(-\frac{1}{2}\left(A_{ij}\hat{\theta}_i\hat{\theta}_j + B_{i\alpha}\hat{\theta}_i X_\alpha + C_{\alpha\beta}X_\alpha X_\beta\right)\right)\ , \tag{42}$$

where **a**, **b**, **c**, **A**, **B**, and **C** are matrices to be defined later.

We will now rewrite the Landau collision integral, presented in the rest frame as Eq. (1), in the lab frame. Recalling that the rest-frame particle velocities are $\mathbf{v}_r = \beta c\gamma\hat{\boldsymbol{\theta}}$, where $\hat{\boldsymbol{\theta}}$ is the lab frame angle of each particle, and transforming Eq. (1) to the lab frame, we now have

$$\frac{\partial f}{\partial t} = -\frac{2\pi r_0^2 c L_c}{\beta^3\gamma^5}\frac{N}{C}\frac{\partial}{\partial\hat{\theta}_i}\int\left(f\frac{\partial f'}{\partial\hat{\theta}'_j} - f'\frac{\partial f}{\partial\hat{\theta}_j}\right)\frac{u^2\delta_{ij}-u_iu_j}{u^3}\delta\left(\mathbf{X}-\mathbf{X}'\right)d^2X'd^3\hat{\theta}'\ , \quad \mathbf{u}=\hat{\boldsymbol{\theta}}-\hat{\boldsymbol{\theta}}'\ , \tag{43}$$

where we assumed a coasting beam of $N$ particles with a uniform longitudinal density, circulating in a ring, and $\int f\, d^2\mathrm{X}\, d^3\hat{\theta} = 1$. Similar to Eq. (1) $f \equiv f(\mathbf{X},\hat{\boldsymbol{\theta}})$ and $f' \equiv f(\mathbf{X}',\hat{\boldsymbol{\theta}}')$.

## 4. Emittance growth rates

We can now calculate the emittance growth rates using Eqs. (40):

$$\begin{aligned}\frac{d\varepsilon_q}{dt} &= \left\langle \int J_q\frac{\partial f}{\partial t}d^2Xd^3\hat{\theta}\right\rangle_s \\ \frac{d\sigma_p{}^2}{dt} &= \gamma^2\left\langle\int\hat{\theta}_s{}^2\frac{\partial f}{\partial t}d^2Xd^3\hat{\theta}\right\rangle_s\end{aligned}\ , \tag{44}$$

where $\partial f/\partial t$ can be evaluated using the Landau collision integral (43), and $\langle\ \rangle_s \equiv \oint(...)ds/C$ denotes averaging over the ring circumference. After some lengthy derivations, similar to Eqs. (9) - (20), we obtain

$$\frac{d\varepsilon_q}{dt} = \frac{Nr_0^2c}{8\pi^{3/2}\beta^3\gamma^4C\varepsilon_1\varepsilon_2\sigma_p}\left\langle L_c\int a_{il}^{(q)}\exp\left(-\frac{1}{4}\mathbf{u}^T\mathbf{A}\mathbf{u}\right)\left(\frac{u^2\delta_{il}-3u_iu_l}{u^3}\right)d^3u\right\rangle_s\ , \tag{45}$$

$$\frac{d\sigma_p^2}{dt} = \frac{Nr_0^2c}{8\pi^{3/2}\beta^3\gamma^4C\varepsilon_1\varepsilon_2\sigma_p}\left\langle L_c\int\gamma^2\exp\left(-\frac{1}{4}\mathbf{u}^T\mathbf{A}\mathbf{u}\right)\left(\frac{u^2-3u_3^2}{u^3}\right)d^3u\right\rangle_s\ , \tag{46}$$

where $\mathbf{u}=\hat{\boldsymbol{\theta}}-\hat{\boldsymbol{\theta}}'$, $\hat{\boldsymbol{\theta}} = \begin{bmatrix}\theta_x & \theta_y & \hat{\theta}_s\end{bmatrix}^T$, and $(q)$ = 1, 2. One can notice that only the matrices $\mathbf{a}^{(q)}$ and **A** need to be determined. Using Eqs. (32) and (42) we can write

$$\mathbf{A}=\begin{bmatrix} \Xi_{22} & \Xi_{24} & \gamma\Xi_{25} \\ \Xi_{24} & \Xi_{44} & \gamma\Xi_{45} \\ \gamma\Xi_{25} & \gamma\Xi_{45} & \gamma^2\Xi_{55} \end{bmatrix} . \tag{47}$$

Similarly, Eqs. (39) and (41) lead to

$$J_q=\frac{1}{2}\left(\mathbf{x}^T-\mathbf{D}^T\theta_s\right)\tilde{\mathbf{V}}^{(q)}\left(\mathbf{x}-\mathbf{D}\theta_s\right)=\frac{1}{2}\left(\mathbf{x}^T\tilde{\mathbf{V}}^{(q)}\mathbf{x}+2\mathbf{x}^T\tilde{\mathbf{V}}^{(q)}\mathbf{D}\theta_s+\theta_s{}^2\mathbf{D}^T\tilde{\mathbf{V}}^{(q)}\mathbf{D}\right)$$

$$=\frac{1}{2}\begin{bmatrix} x \\ \hat{\theta}_x \\ y \\ \hat{\theta}_y \\ \hat{\theta}_s \end{bmatrix}^T \begin{bmatrix} \tilde{V}_{11}^{(q)} & \tilde{V}_{12}^{(q)} & \tilde{V}_{13}^{(q)} & \tilde{V}_{14}^{(q)} & \gamma\tilde{V}_{15}^{(q)} \\ \tilde{V}_{12}^{(q)} & \tilde{V}_{22}^{(q)} & \tilde{V}_{23}^{(q)} & \tilde{V}_{24}^{(q)} & \gamma\tilde{V}_{25}^{(q)} \\ \tilde{V}_{13}^{(q)} & \tilde{V}_{23}^{(q)} & \tilde{V}_{33}^{(q)} & \tilde{V}_{34}^{(q)} & \gamma\tilde{V}_{35}^{(q)} \\ \tilde{V}_{14}^{(q)} & \tilde{V}_{24}^{(q)} & \tilde{V}_{34}^{(q)} & \tilde{V}_{44}^{(q)} & \gamma\tilde{V}_{45}^{(q)} \\ \gamma\tilde{V}_{15}^{(q)} & \gamma\tilde{V}_{25}^{(q)} & \gamma\tilde{V}_{35}^{(q)} & \gamma\tilde{V}_{45}^{(q)} & \gamma^2\mathbf{D}^T\tilde{\mathbf{V}}^{(q)}\mathbf{D} \end{bmatrix} \begin{bmatrix} x \\ \hat{\theta}_x \\ y \\ \hat{\theta}_y \\ \hat{\theta}_s \end{bmatrix}, \quad \begin{bmatrix} \tilde{V}_{15}^{(q)} \\ \tilde{V}_{25}^{(q)} \\ \tilde{V}_{35}^{(q)} \\ \tilde{V}_{45}^{(q)} \end{bmatrix}=\tilde{\mathbf{V}}^{(q)}\mathbf{D} , \tag{48}$$

which results in

$$\mathbf{a}^{(q)}=\frac{1}{2}\begin{bmatrix} \tilde{V}_{22}^{(q)} & \tilde{V}_{24}^{(q)} & \gamma\tilde{V}_{25}^{(q)} \\ \tilde{V}_{24}^{(q)} & \tilde{V}_{44}^{(q)} & \gamma\tilde{V}_{45}^{(q)} \\ \gamma\tilde{V}_{25}^{(q)} & \gamma\tilde{V}_{45}^{(q)} & \gamma^2\mathbf{D}^T\tilde{\mathbf{V}}^{(q)}\mathbf{D} \end{bmatrix}, \quad \mathbf{a}^{(s)}=\begin{bmatrix} 0 & 0 & 0 \\ 0 & 0 & 0 \\ 0 & 0 & \gamma^2 \end{bmatrix}, \tag{49}$$

where $q=1,2$. Notice that matrices $\mathbf{a}^{(q)}$ have the units of *length*, while $\mathbf{a}^{(s)}$ is unitless. We can now introduce the 3x3 matrix **T**, which performs a rotation in the velocity space and diagonalizes the matrix **A** so that,

$$\begin{gathered} \mathbf{T}^T\mathbf{A}\mathbf{T}\equiv T_{ij}A_{jk}T_{kl}=\mathrm{diag}(A_1,A_2,A_3)\equiv\mathrm{diag}(1/\sigma_1{}^2,1/\sigma_2{}^2,1/\sigma_3{}^2)\equiv\mathbf{A}' , \\ T_{ik}T_{jk}=\delta_{ij} , \quad \mathbf{T}^T\mathbf{T}=\mathbf{I} \end{gathered} . \tag{50}$$

Using this diagonalized matrix $\mathbf{A}'$ we obtain:

$$\begin{aligned} \frac{d\varepsilon_q}{dt} &= \frac{r_0^2 cN}{8\pi^{3/2}\beta^3\gamma^4 C\varepsilon_1\varepsilon_2\sigma_p}\left\langle L_c\int a_{il}^{(q)}\exp\left(-\frac{1}{4}\mathbf{u}^T\mathbf{A}'\mathbf{u}\right)\left(\frac{u^2\delta_{il}-3T_{in}T_{lm}u_n u_m}{u^3}\right)d^3u\right\rangle_s \\ &= \frac{r_0^2 cN}{8\pi^{3/2}\beta^3\gamma^4 C\varepsilon_1\varepsilon_2\sigma_p}\left\langle L_c\int\exp\left(-\frac{1}{4}\mathbf{u}^T\mathbf{A}'\mathbf{u}\right)\left(\frac{a_{ii}^{(q)}}{u}-3\left[\mathbf{T}^T\mathbf{a}^{(q)}\mathbf{T}\right]_{nm}\frac{u_n u_m}{u^3}\right)d^3u\right\rangle_s \end{aligned} . \tag{51}$$

Considering that only terms with even powers contribute to the integral we obtain

$$\frac{d\varepsilon_q}{dt}=\frac{Nr_0^2 c}{8\pi^{3/2}\beta^3\gamma^4 C\varepsilon_1\varepsilon_2\sigma_p}\left\langle L_c\left(\sum_{i=1}^{3}a_{ii}^{(q)}\int \mathrm{e}^{-\frac{1}{4}\mathbf{u}^T\mathbf{A}'\mathbf{u}}\frac{d^3u}{u}-3\sum_{i=1}^{3}\left[\mathbf{T}^T\mathbf{a}^{(q)}\mathbf{T}\right]_{ii}\int\frac{u_i{}^2}{u^3}d^3u\right)\right\rangle_s . \tag{52}$$

Using the techniques similar to Eqs. (11) and (12) we finally obtain

$$\frac{d\varepsilon_q}{dt}=\frac{Nr_0^2c}{3\sqrt{\pi}\beta^3\gamma^4 C\varepsilon_1\varepsilon_2\sigma_p}\Big\langle L_c\sigma_1\sigma_2\sigma_3\Big(\sigma_1^{\ 2}\,\mathrm{R_D}\left(\sigma_2^{\ 2},\sigma_3^{\ 2},\sigma_1^{\ 2}\right)\left(\mathrm{Sp}\left(\mathbf{a}^{(q)}\right)-3\left[\mathbf{T}^T\mathbf{a}^{(q)}\mathbf{T}\right]_{11}\right)$$
$$+\sigma_2^{\ 2}\,\mathrm{R_D}\left(\sigma_3^{\ 2},\sigma_1^{\ 2},\sigma_2^{\ 2}\right)\left(\mathrm{Sp}\left(\mathbf{a}^{(q)}\right)-3\left[\mathbf{T}^T\mathbf{a}^{(q)}\mathbf{T}\right]_{22}\right)+\sigma_3^{\ 2}\,\mathrm{R_D}\left(\sigma_1^{\ 2},\sigma_2^{\ 2},\sigma_3^{\ 2}\right)\left(\mathrm{Sp}\left(\mathbf{a}^{(q)}\right)-3\left[\mathbf{T}^T\mathbf{a}^{(q)}\mathbf{T}\right]_{33}\right)\Big)\Big\rangle_s\ , \qquad (53)$$

where the function Sp(**a**) denotes the matrix trace. The momentum spread growth rate is

$$\frac{d\sigma_p^2}{dt}=\frac{Nr_0^2c}{3\sqrt{\pi}\beta^3\gamma^4 C\varepsilon_1\varepsilon_2\sigma_p}\Big\langle L_c\sigma_1\sigma_2\sigma_3\Big(\sigma_1^{\ 2}\,\mathrm{R_D}\left(\sigma_2^{\ 2},\sigma_3^{\ 2},\sigma_1^{\ 2}\right)\left(\mathrm{Sp}\left(\mathbf{a}^{(s)}\right)-3\left[\mathbf{T}^T\mathbf{a}^{(s)}\mathbf{T}\right]_{11}\right)$$
$$+\sigma_2^{\ 2}\,\mathrm{R_D}\left(\sigma_3^{\ 2},\sigma_1^{\ 2},\sigma_2^{\ 2}\right)\left(\mathrm{Sp}\left(\mathbf{a}^{(s)}\right)-3\left[\mathbf{T}^T\mathbf{a}^{(s)}\mathbf{T}\right]_{22}\right)+\sigma_3^{\ 2}\,\mathrm{R_D}\left(\sigma_1^{\ 2},\sigma_2^{\ 2},\sigma_3^{\ 2}\right)\left(\mathrm{Sp}\left(\mathbf{a}^{(s)}\right)-3\left[\mathbf{T}^T\mathbf{a}^{(s)}\mathbf{T}\right]_{33}\right)\Big)\Big\rangle_s\ . \qquad (54)$$

The Eqs. (53) and (54) are the main result of this paper. In typical rings, the synchrotron tune is much smaller than 1. This results in good decoupling between transverse and longitudinal motions and allows one to use Eqs. (53) and (54) for IBS calculations in a bunched beam. That requires two substitutions. The first replacement is:

$$C\rightarrow 2\sqrt{\pi}\sigma_s\,, \qquad (55)$$

where $\sigma_s$ is the rms bunch length. This substitution is applicable only to the denominators of Eqs. (53) and (54). The second substitution ($3\rightarrow 6$) needs to be done in the denominator of Eq. (54). It takes into account that the momentum spread in a bunched beam grows at half the rate due to the redistribution of energy between potential and kinetic energies of the synchrotron motion, which implies linear focusing in the longitudinal plane.

Finally, the Coulomb logarithm, $L_c$, is calculated similarly to the plasma case, Eq. (22), with the following correction, affecting the value of maximum impact parameter

$$\rho_{\min}=r_0c^2/\,\mathrm{Sp}(\mathbf{\Sigma}_{\mathrm{v}})$$
$$\rho_{\max}=\min\left(\sigma_{\min},\gamma\sigma_s,\sqrt{\frac{\mathrm{Sp}(\mathbf{\Sigma}_{\mathrm{v}})}{4\pi n r_0 c^2}}\right)\ , \qquad (56)$$

where $\mathbf{\Sigma}_{\mathrm{v}}=\left(\gamma\beta c\right)^2\mathbf{A}^{-1}$, and $\mathbf{A}$ is given by Eq. (47), and $\sigma_{\min}$ is the smaller of the two transverse rms beam sizes at a given location in a ring so that $\sigma_{\min}^{\ 2}=\left(\Sigma_{11}+\Sigma_{33}-\sqrt{\left(\Sigma_{11}-\Sigma_{33}\right)^2+4\Sigma_{13}^2}\right)/\,2$ with $\mathbf{\Sigma}=\mathbf{\Xi}_{tot}^{\ -1}$.

## 5. IBS in Special Cases

To test the above result, we produce calculations for the uncoupled case in the smooth lattice approximation ($\beta_{x,y}=const$) in the absence of dispersion. Then, the distribution function is

$$f=\frac{\gamma}{4\pi^2\sqrt{2\pi}\varepsilon_x\varepsilon_y\sigma_p}\exp\left(-\frac{1}{2}\left(\frac{x^2}{\varepsilon_x\beta_x}+\frac{\beta_x\hat{\theta}_1^{\,2}}{\varepsilon_x}+\frac{y^2}{\varepsilon_y\beta_y}+\frac{\beta_y\hat{\theta}_2^{\,2}}{\varepsilon_y}+\frac{\gamma^2\hat{\theta}_3^{\,2}}{\sigma_p^{\,2}}\right)\right). \tag{57}$$

This yields for the matrix **A**, already diagonalized, thus $\mathbf{T}=\mathrm{diag}(1,1,1)$:

$$\mathbf{A}=\begin{bmatrix}\beta_x/\varepsilon_x & 0 & 0\\ 0 & \beta_y/\varepsilon_y & 0\\ 0 & 0 & \gamma^2/\sigma_p^{\,2}\end{bmatrix}. \tag{58}$$

The beam emittances are

$$\varepsilon_x=\frac{1}{2}\left(\frac{x^2}{\beta_x}+\beta_x\hat{\theta}_1^{\,2}\right),\quad \varepsilon_y=\frac{1}{2}\left(\frac{x^2}{\beta_y}+\beta_y\hat{\theta}_2^{\,2}\right),\quad \sigma_p^2=\gamma^2\hat{\theta}_3^{\,2}\ . \tag{59}$$

Consequently, the matrix **a** is

$$\mathbf{a}^{(x)}=\frac{1}{2}\begin{bmatrix}\beta_x & 0 & 0\\ 0 & 0 & 0\\ 0 & 0 & 0\end{bmatrix},\quad \mathbf{a}^{(y)}=\frac{1}{2}\begin{bmatrix}0 & 0 & 0\\ 0 & \beta_y & 0\\ 0 & 0 & 0\end{bmatrix},\quad \mathbf{a}^{(s)}=\begin{bmatrix}0 & 0 & 0\\ 0 & 0 & 0\\ 0 & 0 & \gamma^2\end{bmatrix}. \tag{60}$$

We can now use Eq. (53) to obtain the emittance (for example the horizontal one) growth rate. We first recall that

$$\sigma_1=\sqrt{\frac{\varepsilon_x}{\beta_x}},\quad \sigma_2=\sqrt{\frac{\varepsilon_y}{\beta_y}},\quad \sigma_3=\frac{\sigma_p}{\gamma}\ . \tag{61}$$

For the horizontal emittance we have: $\mathrm{Sp}\left(\mathbf{a}^{(x)}\right)=\beta_x/2$. Then, we have

$$\begin{aligned}\frac{d\varepsilon^{(x)}}{dt}&=\frac{NL_c r_0^2 c\sigma_1\sigma_2\sigma_3}{3\sqrt{\pi}\beta^3\gamma^4 C\varepsilon_1\varepsilon_2\sigma_p}\left(\sigma_1^{\,2}\,\mathrm{R_D}\left(\sigma_2^{\,2},\sigma_3^{\,2},\sigma_1^{\,2}\right)\left(\frac{\beta_x}{2}-3\frac{\beta_x}{2}\right)\right.\\&\quad\left.+\sigma_2^{\,2}\,\mathrm{R_D}\left(\sigma_3^{\,2},\sigma_1^{\,2},\sigma_2^{\,2}\right)\left(\frac{\beta_x}{2}-3\cdot 0\right)+\sigma_3^{\,2}\,\mathrm{R_D}\left(\sigma_1^{\,2},\sigma_2^{\,2},\sigma_3^{\,2}\right)\left(\frac{\beta_x}{2}-3\cdot 0\right)\right)\\&=\frac{NL_c r_0^2 c\sigma_1\sigma_2\sigma_3}{3\sqrt{\pi}\beta^3\gamma^4 C\varepsilon_1\varepsilon_2\sigma_p}\frac{\beta_x}{2}\left(\sigma_2^{\,2}\,\mathrm{R_D}\left(\sigma_3^{\,2},\sigma_1^{\,2},\sigma_2^{\,2}\right)+\sigma_3^{\,2}\,\mathrm{R_D}\left(\sigma_1^{\,2},\sigma_2^{\,2},\sigma_3^{\,2}\right)-2\sigma_1^{\,2}\,\mathrm{R_D}\left(\sigma_2^{\,2},\sigma_3^{\,2},\sigma_1^{\,2}\right)\right)\end{aligned} \tag{62}$$

Recalling Eq. (17) we obtain

$$\begin{aligned}\frac{d\varepsilon^{(x)}}{dt}&=\frac{NL_c r_0^2 c\sigma_1\sigma_2\sigma_3}{3\sqrt{\pi}\beta^3\gamma^4 C\varepsilon_1\varepsilon_2\sigma_p}\frac{\beta_x}{2}\frac{3\pi}{\sqrt{2}\sqrt{\sigma_1^{\,2}+\sigma_2^{\,2}+\sigma_3^{\,2}}}\Psi\left(\sigma_1,\sigma_2,\sigma_3\right)\\&=\frac{\sqrt{\pi}NL_c r_0^2\sigma_1\sigma_2\sigma_3\beta_x}{2\sqrt{2}\beta^3\gamma^4 C\varepsilon_1\varepsilon_2\sigma_p}\frac{\Psi\left(\sigma_1,\sigma_2,\sigma_3\right)}{\sqrt{\sigma_1^{\,2}+\sigma_2^{\,2}+\sigma_3^{\,2}}}\end{aligned}, \tag{63}$$

which coincides with the result previously obtained in Refs. [6, 7]. The results for other degrees of freedom are similarly identical to the previously obtained for uncoupled optics.

## Conclusions

In conclusion, we would like to point out the applicability conditions of the considered IBS model. First, similar to the uncoupled case, the considered model implies that the distribution function stays Gaussian in the process of its evolution. In practical terms this approximation is quite good. However, if the rms velocities of different modes (mode emittances) are significantly different, the non-Gaussian tails will appear. For the case when the mode temperatures are different by many orders of magnitude these tails are produced by single Coulomb scattering events (the so-called Touschek effect) and their effect can be accounted for independently from the scattering in the core. Otherwise, an integro-differential equation is required to describe the combined process [11]. Second, the presented IBS model is applicable in the logarithmic approximation only, *i.e.* the Coulomb logarithm (introduced in Eq. (20)) must be much larger than 1, and the beam is assumed to be non-relativistic in the beam frame.

## Acknowledgments

This manuscript has been authored by Fermi Research Alliance, LLC under Contract No. DE-AC02-07CH11359 with the U.S. Department of Energy Office of Science, Office of High Energy Physics.

## References

1. E.M. Lifshitz and L. P. Pitaevskii, “Physical Kinetics,” v.10, Pergamon Press (1981).
2. A. Piwinski, Proc. 9th Int. Conf. on High Energy Accelerators, p. 405 (1974).
3. J. D. Bjorken and S. K. Mtingwa, Particle Accelerators 13, 115 (1983).
4. L.D. Landau, Phys. Z. Sowjetunion, **10**, 154 (1936).
5. M. N. Rosenbluth, W.M. MacDonald, and D. L. Judd, Phys. Rev. 107, 1 (1957).
6. “Accelerator physics at the Tevatron collider”, editors: V. Lebedev and V. Shiltsev, Springer, 2014.
7. S. Nagaitsev, Phys. Rev. ST Accel. Beams 8, 064403 (2005).
8. K. Kubo and K. Oide, Phys. Rev. ST Accel. Beams 4, 124401 (2001).
9. G. I. Budker and A.N. Skrinsky, Sov. Phys. Usp. 21(4), April 1978.
10. V.A. Lebedev and S. A. Bogacz, “Betatron motion with coupling of horizontal and vertical degrees of freedom”,2010 *JINST* **5** P10010.

11. V. Lebedev, J.S Hangst, N Madsen, and A. Labrador, "Single and multiple intrabeam scattering in a laser cooled beam", NIM-A, v. 391, p. 176 (1997)